\documentclass[10pt,twocolumn,brazil,english,aps,prl]{revtex4-1}
\usepackage{mathptmx}

\newcommand{\norm}[1]{\lVert#1\rVert}
\usepackage[T1]{fontenc}
\usepackage[latin9]{inputenc}
\usepackage{color}
\usepackage{epsfig}
\usepackage{prettyref}
\usepackage{textcomp}
\usepackage{amssymb}
\usepackage{amsmath}
\usepackage{babel}


\makeatother

\begin{document}

\preprint{This line only printed with preprint option}

\title{Energy transfer dynamics and thermalization of two oscillators
interacting via chaos}

\author{M.~A.~Marchiori, Ricardo Fariello and M.~A.~M.~de Aguiar}

\affiliation{Instituto de F\'{\i}sica `Gleb Wataghin', Universidade
Estadual de Campinas (UNICAMP), 13083-970 Campinas, SP, Brazil}

\begin{abstract}

We consider the classical dynamics of two particles moving in harmonic
potential wells and interacting with the same external environment
$H_E$, consisting of $N$ non-interacting chaotic systems. The
parameters are set so that when either particle is separately placed in
contact with the environment, a dissipative behavior is observed. When
both particles are simultaneously in contact with $H_E$ an
indirect coupling between them is observed only if the particles are in
near resonance. We study the equilibrium properties of the system
considering ensemble averages for the case $N=1$ and single trajectory
dynamics for $N$ large. In both cases, the particles and the environment
reach an equilibrium conf\/iguration at long times, but only for large $N$
a temperature can be assigned to the system.

\end{abstract}

\maketitle

\section{Introduction}

Understanding dissipation from a microscopic point of view has become
important to several areas of physics, especially if quantum phenomena
are relevant. In these cases a globally conservative approach for the
system plus its environment is highly desirable, allowing a direct
quantum mechanical description.

The simplest, and perhaps most natural, way to model the environment is
to use an inf\/inite set of harmonic oscillators, representing the normal
modes of a general system in equilibrium weakly perturbed by the system
of interest \cite{caldeira83,caldeira}. The spectral function, which is
related to the distribution of frequencies of the normal modes, can be
chosen to model several types of thermal baths
\cite{fisher85,weiss93,hedegard87}. Other representations of the
environment have also been explored, from spin systems
\cite{prokofiev00} (or two-level atoms)\ to chaotic systems
\cite{wilkinson90,berry93,tulio,cohen,cohen1}. The latter is
particularly important to model coupling to small external systems
where the chaotic nature of the trajectories compensates for the small
number of degrees of freedom in the decay of correlation functions.
However, chaos alone does not suf\/f\/ice to simulate a thermal bath,
since small numbers of degrees of freedom always leave a strong
signature in the dynamics through large f\/luctuations in the
observables of the system of interest \cite{marchiori11}. These f\/luctuations
can be washed out by averaging over several realizations of the
dynamics \cite{mm06}.

More recently, the interest has shifted from a single particle
interacting with the environment to two particles independently
connected to it \cite{dua06,dua09,val10}, allowing the study of
interactions mediated by the environment. One
important case is that of two entangled particles subjected to dissipation
and decoherence. In many cases, the total Hamiltonian is symmetric by
the exchange of the particles, although they may still be considered
distinguishable in some applications \cite{dua06}.

In this paper we consider the classical dynamics of two particles moving
in harmonic potentials linearly coupled to a f\/inite chaotic
environment. We study the system behavior as a function of the
frequency of the oscillators. We f\/ind that the behavior of the system
changes radically when the oscillators are in resonance, which is a
necessary condition if the particles are identical, and that even small
deviations from this symmetric state changes dramatically the equilibrium
properties of the system. In particular, we show that, when in resonance,
the particles may exchange energy through the environment while their
energies dissipate. Moreover, the energy stored in their relative motion
is conserved.

We model the environment as a set of $N$ independent quartic systems
(QS), each with 2 degrees of freedom. The QS has a single parameter
that controls the degree of chaos in the dynamics. The particles are
represented by two harmonic oscillators independently coupled to the
set of QS's. As study cases we consider the systems with $N=1$ and
$N=100$, for which we study the statistical properties of the energy
distributions of the environment and of the oscillators. In both cases, these
distributions reach an asymptotic equilibrium, but attempts to def\/ine a
temperature for the system using two basic def\/initions of entropy,
applicable to small systems, succeed only in the case $N=100$. The results
of the simulations are then interpreted in the light of the linear
response theory.

\section{Model}

We consider two harmonic oscillators interacting with an
environment composed by a collection of independent quartic systems.
The Hamiltonian is
\begin{equation}
H=H_{1}+H_{2}+H_{E}+\lambda_{N}H_{I},
\label{eq1}
\end{equation}
where $\lambda_{N}=\lambda/\sqrt{N}$ and
\begin{eqnarray}
H_{i} & = &
\frac{p^{2}_{i}}{2m_{i}}+\frac{m_{i}\omega_{i}^{2}q_{i}^{2}}{2}, \quad i = 1,2,
\label{oscillator}\\
H_{E} & = & \sum_{n=1}^{N}\left[\frac{p_{x_{n}}^{2}+p_{y_{n}}^{2}}{2}+
\frac{a}{4}(x_{n}^{4}+y_{n}^{4})+\frac{x_{n}^{2}y_{n}^{2}}{2}\right]
\equiv \sum_{n=1}^{N} H_{QS}^{(n)},
\label{quartic system}\\
H_{I} & = & \sum_{n=1}^{N}(q_1 + q_2)\, x_{n}.
\label{coupling}
\end{eqnarray}
The total energy is conserved and the two oscillators interact only via
the environment.

In our simulations, we used $\lambda=0.01$. The
parameter $a$ in $H_{QS}$ controls the dynamical regime of the quartic
systems in the environment, ranging from integrable (for the special
values $a=1.0$ and $a=0.33$ \cite{joy93})\ to chaotic ($a\rightarrow 0$).
In the work, we used $a=0.01$ or $a=0.1$, which correspond to regimes
where the QS is mostly chaotic. For $a=0.1$ the largest Lyapunov
exponent of an isolated QS is $\lambda_L=0.166$ for energy $E_{QS} = 0.1$. For
$a=0.01$ we obtained $\lambda_L=0.121$ for $E_{QS} = 0.01$. Because of the
scaling properties of the QS, the Lyapunov exponent at
other energies can be calculated using $\lambda_L(E) = (E/E_0)^{1/4}
\lambda_L(E_0)$.

As the number of degrees of freedom is f\/inite, the parameter $N$
assumes a prominent role and we investigate its inf\/luence in
the dynamic behavior of the oscillators. It has been recently shown
\cite{marchiori11} that, for $N$ suf\/f\/iciently large, such f\/inite chaotic
environment can simulate the action of an inf\/inite thermal
reservoir. Here the environment also acts as a medium connecting the two
harmonic oscillators.

In order to highlight the interaction between the oscillators, one of them
is initialized with energy $E_{1} > 0$ while the other is set at rest with
$E_{2}=0$. For the environment we def\/ine the initial conditions using
the \textit{pseudo canonical} distribution \cite{marchiori11}
\begin{equation}
\rho=\frac{1}{Z}\prod_{n=1}^{N}\delta(H_{QS}^{(n)}-E_{QS}^{(n)}),
\label{eqdist}
\end{equation}
where the energy $E_{QS}^{(n)}$ of each QS is randomly chosen from the exponential
probability distribution $\exp{(-E/\Bar{E}_{QS})}/\Bar{E}_{QS}$. The value
of $\Bar{E}_{QS}$ plays the role of an initial ``temperature'' for the
environment. In the
special case of $N=1$ the energy $E_{QS}$ is f\/ixed to $\Bar{E}_{QS}$.

In what follows we will study the system \eqref{eq1} for only two
relevant environment sizes: the ``microscopic'' ($N=1$)\ and
``macroscopic'' ($N=100$)\ cases. As pointed out in \cite{marchiori11}, the
dynamical behavior of the system becomes $N$-independent for $N$
suf\/f\/iciently large. For the parameters values used in this paper, the large $N$
limit is already reached for $N=100$. The case $N = 1$ is a
natural extension of the work presented in \cite{mm06}, where the authors
considered the interaction between a harmonic oscillator
and a single quartic system using ensemble averages. Thence
we choose the same set of parameters as in \cite{mm06} in our simulations,
allowing us to verify the implications of adding a
second oscillator to the system. For $N = 100$, on the other
hand, we compare our results with the work presented in \cite{marchiori11},
which treated the dynamics of a single oscillator interacting
with large chaotic environments.

In the microscopic case, observables related to the harmonic oscillators,
like the energies $H_1$ or $H_2$, exhibit large f\/luctuations when
coupled to $H_{QS}$. These f\/luctuations can only be washed out
by averaging over large ensembles of realizations of the dynamics. In the
macroscopic case this is not necessary and the results obtained from a
single realization of the dynamics are already representative of the
average behavior. In this case we can speak of an ``ef\/fective
dynamics'', where no averages are performed.

\section{Measures of temperature}

Temperature is a central property in the description of equilibrium and
is properly def\/ined only in the thermodynamic limit of very large
systems. Since the environment def\/ined in Eq.~\eqref{quartic system} is far from
this limit for $N=1$ and $N=100$, dif\/ferent possibilities arise. One
natural def\/inition comes from the equipartition theorem
\cite{rt1918}
\begin{equation}
\Big\langle z_m \frac{\partial H}{\partial z_n}\Big\rangle=
\delta_{mn} \tau_E,
\label{equipartition}
\end{equation}
where $z_n$ denotes the coordinates or momenta of the system and
$\delta_{mn}$ is the Kronecker delta. The constant $\tau_E$ is identif\/ied
with $k_B T$ when the system is in contact with a thermal reservoir. The
subscript in $\tau_E$ emphasizes the explicit use of the equipartition
theorem. Eqs.~\eqref{eq1} and \eqref{equipartition} predict that
the fraction of the total energy within each subsystem in equilibrium
should be $\langle E_1 \rangle = \langle E_2 \rangle = \tau_E =
2/3 \langle
E_{QS} \rangle$. We can also write the system's temperature as a
function of the number of quartic systems as
\begin{equation}
E_T = E_{1}(0) + E_{2}(0) + E^{T}_{QS}(0) = 2\tau_E + \frac{3}{2}N \tau_E,
\label{kt}
\end{equation}
in which $E^{T}_{QS}(0)$ is the environment's total energy at $t=0$.

Eq.~\eqref{kt} can also be derived from the thermodynamic relation
\cite{be91}
\begin{equation}
\tau_E^{-1} = \frac{\partial\ln{\Gamma}}{\partial E_T},
\label{taue}
\end{equation}
where
\begin{equation}
\Gamma(E_T) = \int \Theta(E_T-H) \mspace{1.mu} dp\mspace{1.mu}dq
\label{gamma}
\end{equation}
and the integral is taken over the entire phase space of the system.
For the Hamiltonian in Eq.~\eqref{eq1} we obtain
\begin{equation}
\Gamma = c\, E_T^{3N_f/4+2},
\label{gamma1}
\end{equation}
where $N_f = 2N$ is the number of degrees of freedom of the environment and
$c$ represents a constant depending on the system parameters.

Notice that
\begin{equation}
\ln{\Gamma} = \ln{c} + (3N_f/4+2) \ln{E_T}
\label{ln-gamma}
\end{equation}
plays the role of entropy. The usual entropy, on the other hand, is given by
\begin{eqnarray}
S(E) &=& \ln{\left[\int \delta(E - H) \mspace{1.mu} dp\mspace{1.mu}dq \right]}
  =  \ln{\frac{\partial\Gamma}{\partial E}} \nonumber \\
 & = & \ln{c} + (3N_f/4+1) \ln{E} + \ln{(3N_f/4+2)}.
\label{sent}
\end{eqnarray}
Def\/ining $\tau_S^{-1} = \partial S(E_T)/\partial E_T$ we obtain
\begin{equation}
E_T = \tau_S + \frac{3}{2}N \tau_S,
\label{taus}
\end{equation}
which agrees with \eqref{kt} in the limit of large $N$.

\begin{figure*}[t]
\begin{center}
\includegraphics[scale=1.05]{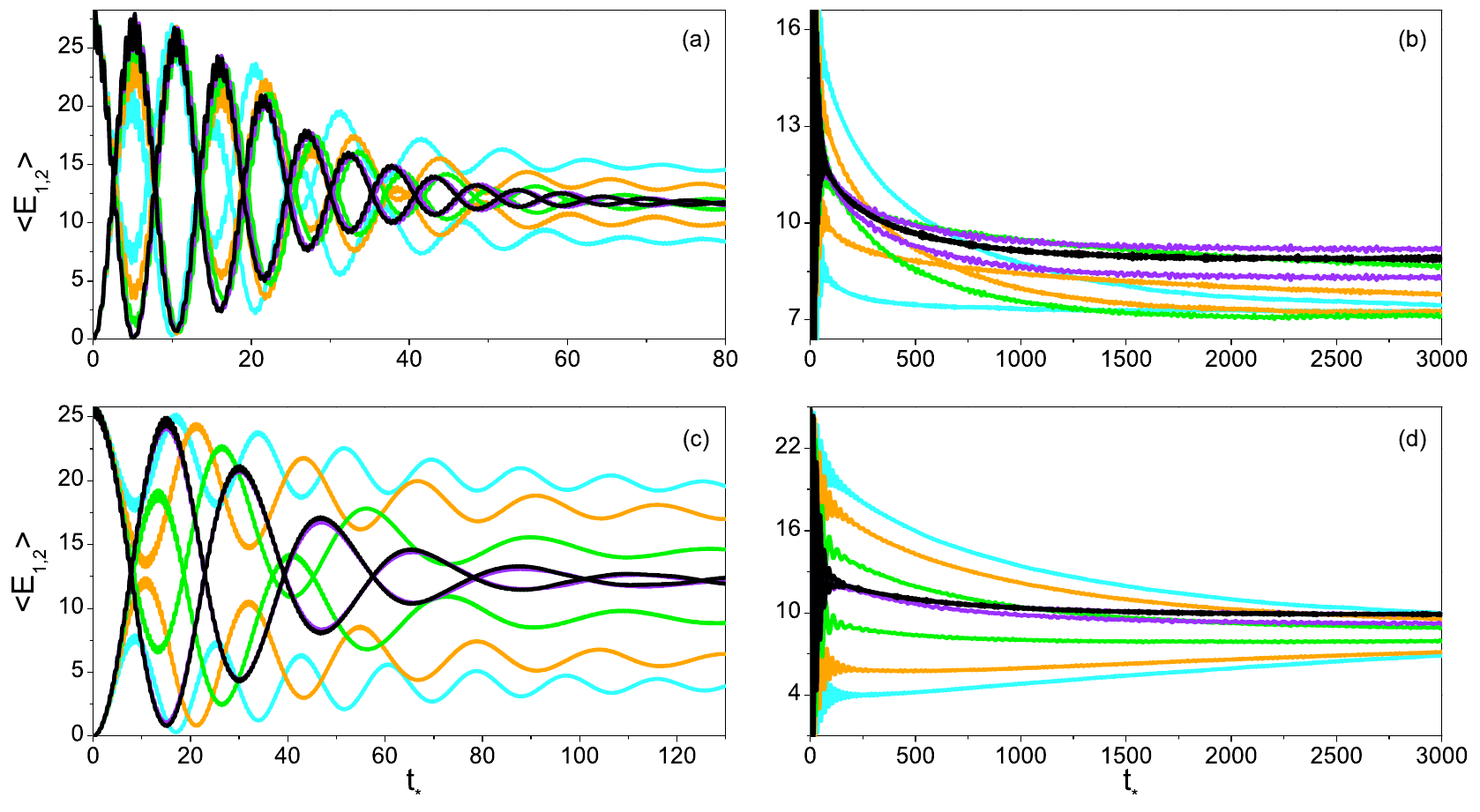} 
\end{center}
\caption{(Color online)\ Short- and long-time energy dynamics of the
oscillators at various values of $k = 0$ (black), $1$ (purple), $6$ (green),
$11$ (orange), $16$ (cyan)\ for energies of $E_{QS \mspace{-1.7mu}}(0) = 0.1$
(panels (a)\ and (b))\ and $E_{QS\mspace{-1.7mu}}(0) = 5$ (panels (c)\ and
(d)). The averages were computed using an ensemble of $40000$ initial
conditions. Colors change gradually from light gray to black as the $k$ value
decreases, if printed in black and white.}
\label{fig1}
\end{figure*}

The two temperatures, $\tau_E$ and $\tau_S$, can be calculated
numerically in a number of ways. According to \cite{ru97}, $\tau_S$ may
be calculated from the dynamics using the expression \cite{ru97,jep00}
\begin{equation}\label{Req}
\tau_S^d =  {\langle\Phi(H)\rangle}^{-1},
\end{equation}
where $\tau_{S}^d/k_B$ is the so-called Rugh's temperature and
\begin{equation}
\Phi(H)\equiv\text{div}\mspace{4mu}\frac{\nabla H}{\norm{\nabla
H}{}^2}
\end{equation}	
with $\nabla$ denoting the gradient in phase space. This
formula assumes that the energy, $H=E$, is the only isolating integral.
On the other hand, $\tau_E$ can be estimated as a dynamical average, as
in the left hand side of Eq.~\eqref{equipartition}. We use the superscript
$d$ to emphasize that these temperatures are obtained from the dynamics,
i.e., from the trajectories.

Finally, the temperature can be calculated from f\/itting the energy distribution
$p(E)$, which is the probability of f\/inding one QS with energy $E$
when the system is in equilibrium, as the Boltzmann exponential
$p_B(E) \sim \exp{\mspace{-1mu}(-E/\tau_B)}$. Another possibility
is to numerically calculate the distribution of
momentum and f\/it it with the Maxwellian prof\/ile $p_M(p)\sim
\exp{\mspace{-1mu}(-p^2/(2m\tau_M))}$. These distributions will be
obtained from the numerical data for the oscillators and will be used to
check the predictions arising from Eqs.~\eqref{kt} and \eqref{taus}. 

In the limit of large systems, we expect all these measures to approach
the same equilibrium value; however, for small number of degrees of
freedom, they are likely to dif\/fer \cite{vb98}. We will use two basic
def\/initions of temperature, and the dif\/ferent ways to calculate
them, to assess the equilibrium properties of our model system.

\section{Dynamics via ensemble average ($N=1$)}

In this section, we investigate the approach to equilibrium and
equipartition of energy for the system described by the
Hamiltonian \eqref{eq1} with only one quartic system, which means $N=1$. We examine,
particularly, the energy transfer dynamics of this system for cases in
which the coupling between the subsystems is weak and the harmonic oscillators
are near-to-resonance. We test for equilibration by comparing the calculated
distributions of energy and momentum with appropriate equilibrium
distributions. We also analyze and compare the dynamical behavior of $\tau_{E}^{d}$
with that of $\tau_{S}^{d}$, concentrating on the issue of energy equipartition
between the energy stored in the oscillators and in the quartic system
at large times.

Our approach makes use of averaging over large ensembles of dif\/ferent
initial chaotic conf\/igurations that have a common f\/ixed energy
shell where $H_{QS} = E_{QS}$. The initial conditions for the two
harmonic oscillators are $q_1=0$, $p_1=\sqrt{\smash[b]{2m_1
E_{1\mspace{-2mu}}(0)}}$, $q_2=0$, and $p_2=\sqrt{\smash[b]{2m_2
E_{2\mspace{-2mu}}(0)}}$. The initial data used for the phase space
variables $x,y,p_x,p_y$ originate from points
along a single trajectory of the uncoupled chaotic quartic system,
rather than from random starting points in an energy shell.

To solve numerically the equations of motion, we used the fourth-order
Runge-Kutta method (detailed for example in \cite{gg89}). The integration
time step length was set to ensure energy conservation to
within $1\%$ for each individual trajectory. The ensemble average value of
an observable is calculated as the mean of its estimates generated by
propagating initial conditions in the ensemble. Throughout this section, we
set $m_1 = m_2 = 10$ and $a=0.1$~\cite{mm06}.

\begin{figure}[t]
\begin{center}
\includegraphics[scale=1.0]{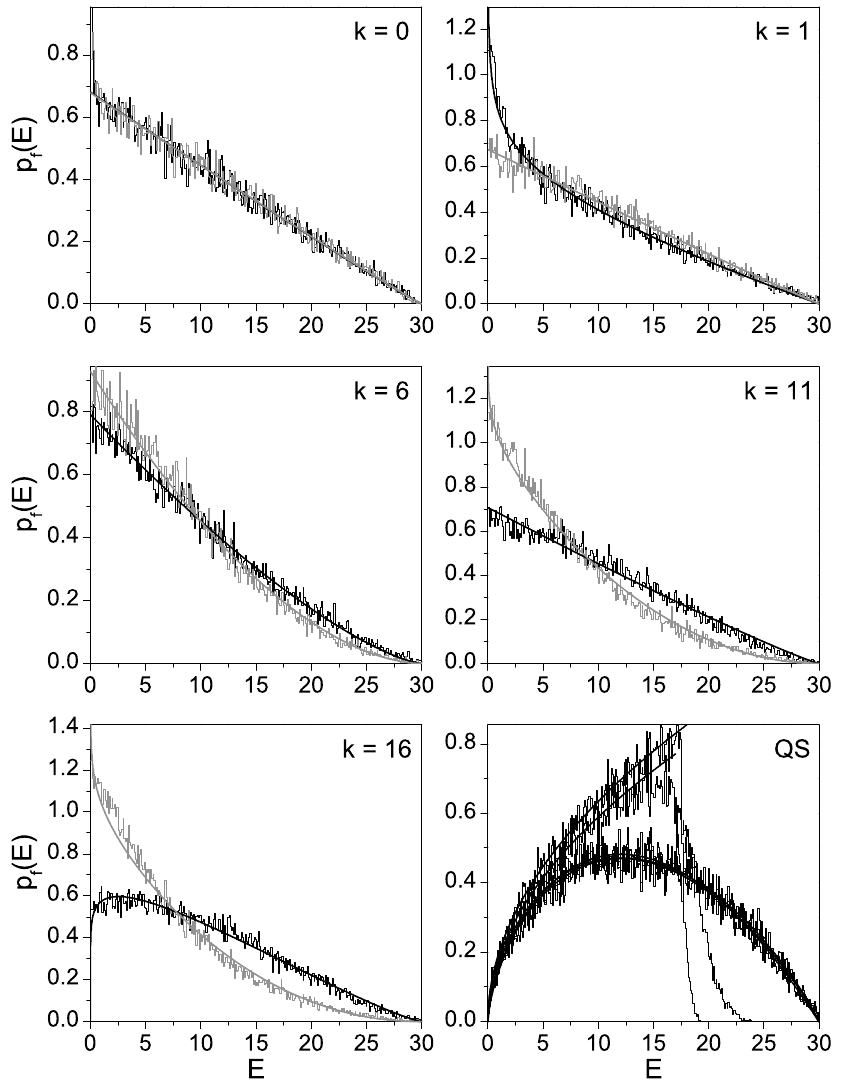}
\end{center}
\caption{Distributions of energy for the three subsystems
at time $t_{\text{f}}$, computed with initial value $E_{QS} = 5$. The
ragged black curves in each $k$-labeled panel are for oscillator 1; the
gray curves are for oscillator 2, and the smooth curves are plotted
according to Eq.~\eqref{fit}. The bottom-right panel displays results for
QS, with $k$ increasing from top curve to bottom curve.}
\label{fig2}
\end{figure}

The left panels in Fig.~\ref{fig1} present the short-time energy transfer
dynamics between the two harmonic oscillators. Their average energies are
plotted versus dimensionless time $t_* = \omega_1 t/2\pi$ for the initial
energies $E_{QS} = 0.1$ (panel (a))\ and $E_{QS} = 5$ (panel (c)), and
frequencies $\omega_2$ in a small deviation from the frequency $\omega_1 =
0.0125$: $\omega_{2} = \omega_1 + k\,\Delta$, with f\/ixed value $\Delta =
0.0000375$ for $k=\{0,1,6,11,16\}$. The initial energies of the
oscillators are chosen as $E_1 = 25$ and $E_2 = 0$. It is readily seen
that the energy curves for $k = 0$ and $1$ are almost identical for both
situations; however the response to $E_{QS}=0.1$ is faster than that of
$E_{QS}=5$. Another important feature is the appreciably less energy
transferred from one oscillator to the other at greater $k$ values.

\begin{figure}[t]
\begin{center}
\includegraphics[scale=1.0]{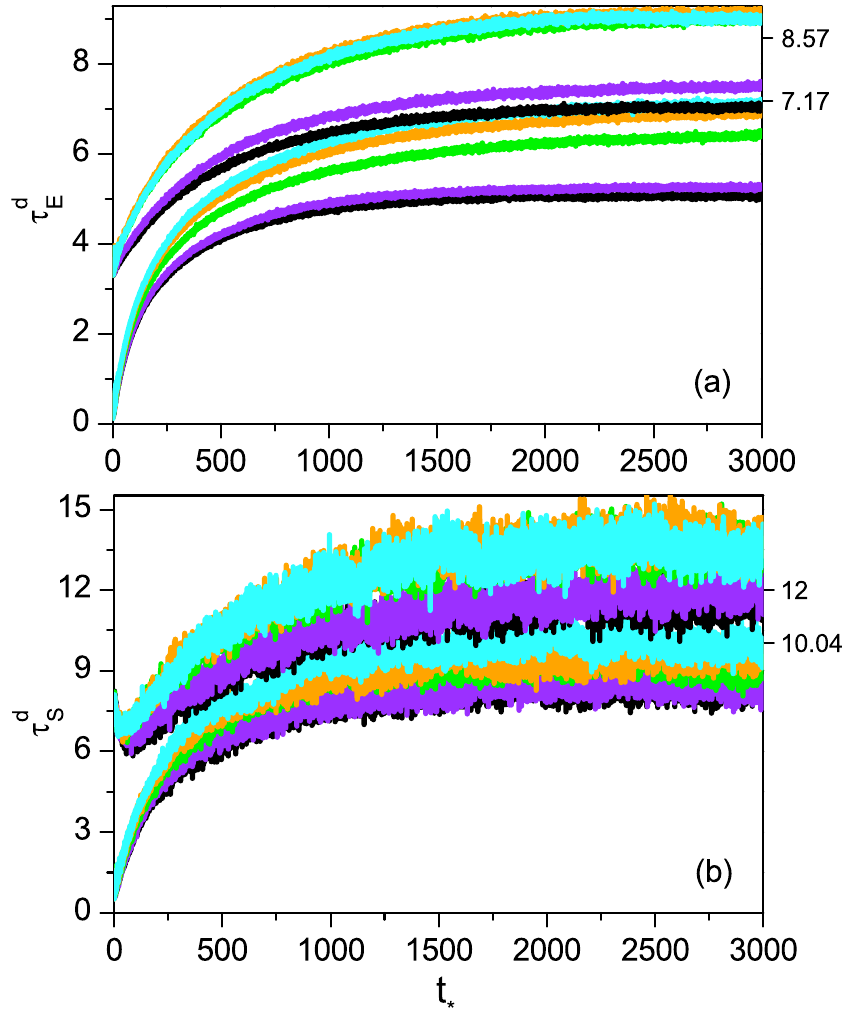}
\end{center}
\caption{(Color online)\ Panel (a): time evolution of $\langle p_x^2 \rangle$
for $E_{QS \mspace{-1.7mu}}(0) = 0.1$ (bottom curves)\ and
$E_{QS \mspace{-1.7mu}}(0) = 5$ (top curves). Panel (b):
time evolution of $\tau_{S}^d$ for $E_{QS \mspace{-1.7mu}}(0) = 0.1$ (bottom
curves)\ and $E_{QS \mspace{-1.7mu}}(0) = 5$ (top curves). Colors and
parameters as in Fig.~\ref{fig1}, except that, for the calculation of
$\tau_{S}^d$, $4000$ initial conditions were used instead of $40000$.
The results for $\langle p_y^2 \rangle$ are nearly identical to those of
$\langle p_x^2 \rangle$.}
\label{fig3}
\end{figure}

The right panels of Fig.~\ref{fig1} in turn capture the long-time behavior
of the average energies of the two oscillators under $E_{QS} = 0.1$ (panel
(b))\ and $E_{QS} = 5$ (panel (d))\ conditions. Except for the case when
the frequencies coincide, the f\/inal energies for the oscillators are
visibly dif\/ferent for each initial state of the system. These panels indicate 
that the equilibration time is very long and might not be reached in practical
situations. Not shown here is the behavior of the chaotic system's 
average energy, which ascends gradually with time and more rapidly 
with increasing $k$, approaching a saturation value. See panel (a)\ in 
Fig.~\ref{fig3} for a plot of $\langle p_x^2 \rangle$ as a function of $t_*$.

\begin{figure*}[t]
\begin{centering}
\includegraphics[scale=0.94]{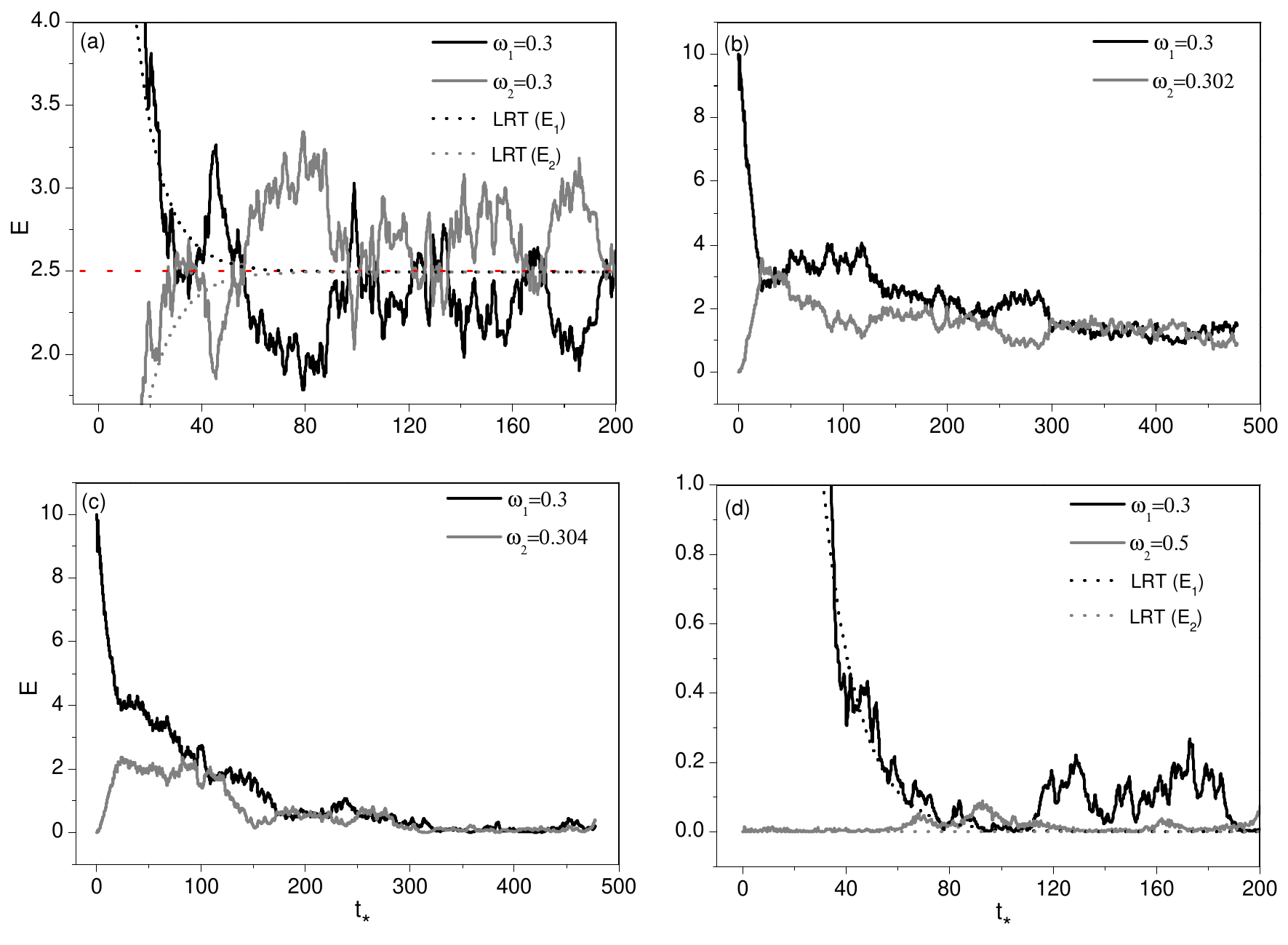}
\end{centering}
\caption{(Color online)\ Energy of the oscillators as a function of the
scaled time $t_*$ for $(\omega_1,\omega_2)=(0.3,0.3)$ (panel (a)),
$(0.3,0.302)$ (panel (b)), $(0.3,0.304)$ (panel (c)), and $(0.3,0.5)$ (panel (d)).
The black and gray solid curves show the numerical results, and black 
and gray dotted curves in panels (a)\ and (d)\ the LRT results calculated
using Eqs.~\eqref{lrt-q1} and \eqref{lrt-q2} with $\gamma \approx 0.0044$
(panel (a))\ and $\gamma \approx 0.0035$ (panel (d)).}
\label{fig4}
\end{figure*}

Fig.~\ref{fig2} exhibits the energy distributions of the two
oscillators for an ensemble of $40000$ initial conditions at time
$t_\text{f}$ --- which corresponds to 3000 periods of the oscillator 1 ---
for the case $E_{QS}(0) = 5$. These distributions are generally
dif\/ferent (except for $k=0$)\ and do not have the Boltzmann form (except
for $k=11$ and $k=16$ for oscillator 2). The f\/ittings to these distributions
are, in majority, of the form
\begin{equation}\label{fit}
p_f(E) = A\mspace{0.4mu}(\langle E_{\mspace{1mu}T
\mspace{-1.7mu}}(t_\text{f})\rangle - E)^B\mspace{-0.1mu}E^C,
\end{equation}
where $\langle E_{\mspace{1mu}T \mspace{-1.7mu}}(t_\text{f})\rangle$ is
the value of the average total system energy at time $t_\text{f}$. In
most cases, all parameters $A$, $B$, and $C$ had non-zero values. One
exception occurs for the chaotic environment for $k$ equal to zero or
one. In such cases, the f\/itting is reduced to
$A\mspace{0.4mu}\sqrt{\smash[b]{E}}$.

For the case of a single harmonic oscillator, the energy distributions
should approach a square root line, as suggested in \cite{mm06}.
However, for $k=11$ and $k=16$, the energy distribution of oscillator 2
can be well f\/it by an exponential, which seems to be an unexpected
transient. Interestingly, the corresponding momentum distribution closely
obey the Maxwellian law $p_M(p)$, which also characterizes approximately
the momentum distributions of the quartic system.
This behavior persists for times up to $t_* =6000$. 
In particular, for $k=11$, the value found of $\tau_{M} = 12.28$
for QS is reasonably close to three-halves of that of $\tau_M = 8.16$ 
for oscillator 2, which in turn is near the corresponding value of $\tau_B = 8.77$. 
This satisf\/ies both equipartition $\langle E_{QS\mspace{-1.7mu}}(t_\text{%
f})  \rangle \approx \tfrac{3}{2}\tau_E$ and, from
energy conservation [cf.~Eq.~\eqref{kt}],
$\tau_E = \tfrac{2}{7}E_{\mspace {1mu}T \mspace{-1.7mu}}(t=0)$. 

We show in Fig.~\ref{fig3} the time evolution of $\tau_{E}^{d}$ associated
with $p_x$ (panel (a))\ and of $\tau_S^d$ (panel (b))\ for all $k$ values
studied. Again, not all curves reach an asymptotic value within the
displayed time window, conf\/irming that dynamical equilibrium was not yet
reached. We compare the asymptotic values with those predicted by
Eqs.~\eqref{kt} and \eqref{taus} for total energies $25.1$ and $30$:
$\tfrac{2}{7}\times 25.1 \approx 7.17$ and $\tfrac{2}{7}\times 30 \approx
8.57$ for panel (a); $\tfrac{2}{5}\times 25.1 = 10.04$ and
$\tfrac{2}{5}\times 30 = 12$ for panel (b). Except for the resonant cases
$k=0$ and $k=1$, $\tau_E^d$ converges to values close to the expected
results of $7.17$ and $8.57$. Also, $\tau_S^d$ converges better to
$12$ than it does to $10.04$ for the resonant cases.

Although both $\tau_{E}^{d}$ and $\tau_S^d$ display very similar time
dependent behavior in practically all cases studied, they disagree with
respect to the mean energy at equilibrium. The numerical results for the
mean oscillators' energies are closer to $\tau_{E}^{d}$ than to
$\tau_{S}^{d}$, which indicates
that the usual def\/inition of entropy [cf.~Eq.~\eqref{sent}] can not be
applied to this system, possibly because of its few degrees of freedom.

\section{Dynamics with a single trajectory ($N=100$)}

In the previous section, we discussed the dynamical behavior and
equilibrium properties of the system described by the Hamiltonian
\eqref{eq1} for $N=1$. The time dependence of any observable, like the
energy of the oscillators, typically displays large f\/luctuations
when calculated for a single initial condition. These f\/luctuations,
which result from the small number of degrees of freedom of the
environment, are drastically reduced when averaged over ensembles of
initial conditions. Even after such average we cannot state that the
environment simulates the action of a thermal reservoir, since the energy
distribution does not always follow a Boltzmann exponential law. As more
QS's are included in the environment, these single trajectory oscillations
decrease. For $N$ suf\/f\/iciently large the time behavior obtained for a
single trajectory becomes similar to that of the ensemble average
and independent of $N$. In this case, we may talk about a potentially {\it
effective} dynamics, where single realizations reproduce the average
behavior.

The indirect interaction between the QS's via the harmonic oscillators
enables the energy to be redistributed among them, leading to a Boltzmann type of
equilibrium distribution for the environment \cite{marchiori11}. When a single 
harmonic oscillator is in contact with a suf\/f\/iciently large and ``cold'' chaotic
environment, most of its energy is transferred to the environment. The results
exhibited in Fig.~\ref{fig1} show that for two resonant
oscillators this is not true: more than 50\% of the initial energy
stored in the oscillators remains in the harmonic modes even for very
long times. In what follows, we explore situations where this symmetry
is broken for $N=100$, for which the environment is already in the
$N$-independent regime. All numerical results in this section are
obtained for a single trajectory and for $m_1 = m_2 = 1$, $a=0.01$~\cite{marchiori11},
and $E_1(0)=10$ and $E_2(0)=0$.

\begin{figure}[h]
\begin{centering}
\includegraphics[scale=1]{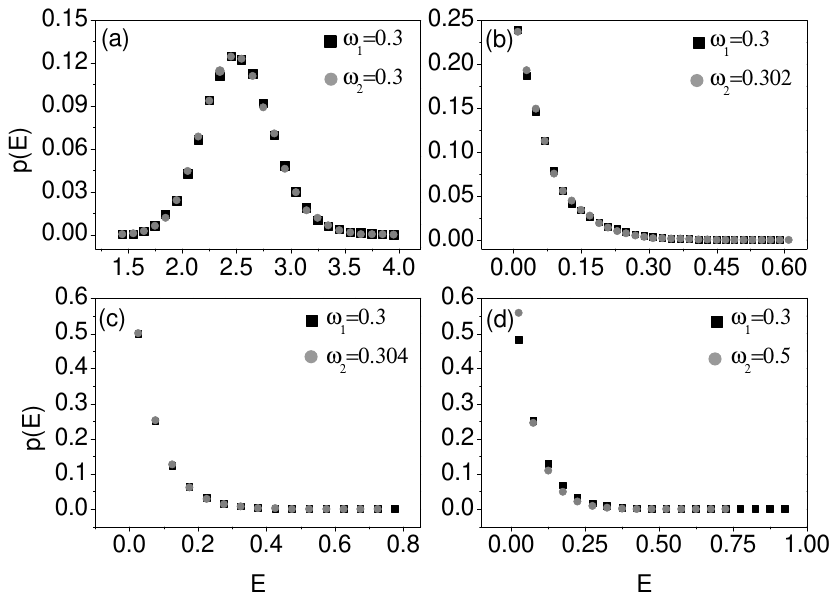}
\end{centering}
\caption{Distributions of energy for oscillator 1 (black squares)\ and 
oscillator 2 (gray circles)\ for the four cases displayed in Fig.~\ref{fig4}. The distribution 
is Gaussian in the resonant case (panel (a), centered in $\Bar{E} = 2.5$), and 
Boltzmann-like in all of\/f-resonant cases: $p_B(E) = \exp{(-E/\Bar{E}_{i})} / \Bar{E}_{i}$ with
$\Bar{E}_{1} \approx 0.072$ and $\Bar{E}_{2} \approx 0.073$
(panel(b)); $\Bar{E}_{1} \approx 0.072$ and $\Bar{E}_{2} \approx
0.073$ (panel(c)); and $\Bar{E}_{1} \approx 0.075$ and
$\Bar{E}_{2}\approx 0.069$ (panel(d)).}
\label{fig5}
\end{figure}

\textit{\textbf{Resonant case}}: Panel (a)\ of Fig.~\ref{fig4} shows the
energy of the two oscillators as a function of time $t_*$ for the resonant
case. Approximately 50\% of $E_1(0)+E_2(0)$ remains with the
oscillators. Notice that the curves displaying $E_1(t_*)$ and
$E_2(t_*)$ are mirror images of each other with respect to the dashed red
horizontal line at $E = 2.5$, which marks the f\/inal mean energy of the two
oscillators. This result is conf\/irmed in panel (a)\ of Fig.~\ref{fig5} where the
energy distributions of the oscillators are Gaussians centered in
$\Bar{E}=2.5$. This is by no means a trivial result and is in
conf\/lict with the equipartition theorem as given in Eq.~\eqref{kt},
which predicts $\Bar{E}_1 = \Bar{E}_2 \approx 0.072$. The condition of
high symmetry associated with the resonance is responsible for this
apparent violation of the equipartition theorem as will be discussed in
the next section.

\textit{\textbf{Non-Resonant case}}: Panels (b)\ to (d)\ in
Fig.~\ref{fig4} show the energy of the oscillators for frequencies
moving away from the resonance. It is still possible to see a f\/low of
energy from one oscillator to the other, although this becomes less
evident as the frequencies become more separated. Also visible is the
increase in the dissipated energy. Fig.~\ref{fig4} also shows how
sensitive is the dynamical behavior of the system with respect to
variations in the frequency of the oscillators. In the quasi-resonant
cases (see panels (b)\ and (c)), where $\omega_1/\omega_2\approx 1$,
the relaxation time is signif\/icantly greater than in the other cases,
allowing energy exchange between the oscillators for a
very long time. However, if the ratio $\omega_1/\omega_2$ deviates more 
considerably from unity (panel (d)), the ``opacity'' of the chaotic medium increases,
culminating in almost independent oscillators that quickly lose all
their energy to the environment. The energy distributions corresponding
to the cases of Fig.~\ref{fig4} are depicted in Fig.~\ref{fig5}, and
show that the f\/inite environment does act as a thermal bath for the two
oscillators, both of which have the same temperature. Moreover, the equipartition
theorem holds true for all three of\/f-resonant cases at long times,
which means that the relation $\Bar{E}_{1} \approx \Bar{E}_{%
2} \approx \tfrac{2}{3}\Bar{E}_{QS}$ is valid. This reinforces the ability
of the f\/inite chaotic environment to promote dissipation and
thermalization \cite{marchiori11}. Nevertheless, the resonance condition
permits an ef\/f\/icient f\/low of energy between the harmonic modes. The aim of
next section is to explain the special resonant case, and show that in this case
the equipartition is still valid.

\section{Linear response theory}

The oscillators obey the dynamics given in the equations
\begin{equation}
\Ddot{q}_i + \omega_{i}^{2}q_i  =  -\frac{\lambda_{N}}{m} \sum_{n=1}^{N}
x_{n}\equiv -\frac{\lambda_{N}}{m}\, X(t) . \label{eqosc}
\end{equation}
As the dynamics of the variables $x_n$ are chaotic and the interaction
among the QS's is of second order in the coupling, each
$x_n$ is approximately independent and we may replace $X(t)$ by its average $\langle
X(t)\rangle$. Applying then linear response theory (LRT)\ \cite{kubo2}, we get
\begin{equation}
\Ddot{q}_i + \omega_{i}^{2}q_i \approx \frac{\lambda^2_N}{m}\int^{t}_{0}
ds\, \phi_{XX}(t-s)(q_{1}(s) + q_{2}(s))\, ,
\end{equation}
where the response function $\phi_{XX}$ is given in Ref.~\cite{marchiori11} by
\begin{equation}
\phi_{XX}(t-s)=\frac{5\mu N}{4}\frac{d}{ds}\delta(t-s)+
\frac{\mu N(t-s)}{4}\frac{d^{2}}{dtds}\delta(t-s)\, ,
\label{response-function}
\end{equation}
where $\mu$ is a parameter that depends on the average energy of the environment. 
After computing the integral we have
\begin{eqnarray}
\Ddot{q}_{1} + \gamma\Dot{q}_{1} + \omega_{1}^{2}q_{1} & =
& -\gamma\Dot{q}_{2},\label{lrt-q1}\\
\Ddot{q}_{2} + \gamma\Dot{q}_{2} + \omega_{2}^{2}q_{2} & =
& -\gamma\Dot{q}_{1} \label{lrt-q2}
\end{eqnarray}
with $\gamma= \frac{3\lambda^2\mu}{8m}$. These equations
cannot be diagonalized unless $\omega_1 = \omega_2$. In this case, we
can def\/ine the new variables
\begin{equation}
Q_{\pm}  =  \frac{ q_{1}\pm q_{2}}{\sqrt{2}},
\label{Q+-}
\end{equation}
associated with center-of-mass and relative coordinates of the
oscillators, and rewrite \eqref{lrt-q1} and \eqref{lrt-q2} as
\begin{eqnarray}\label{new-oscillator}
\Ddot{Q}_{+} + 2\gamma\Dot{Q}_{+}  +  \omega^{2}Q_{+} &=& 0,\\
\Ddot{Q}_{-} + \omega^{2}Q_{-}  &=&  0\label{Q-}.
\end{eqnarray}
The $Q_-$ is then completely decoupled from the environment, and will be
called the \textit{conservative mode}, whose initial energy is $E_1(0)/2$
for the present initial conditions. Analogously, $Q_+$ will be termed
\textit{dissipative mode}. LRT predicts $\gamma \ll \omega$; therefore
$Q_+$ dissipates energy according to $E_{+}(t)\approx
E_{+}(0)\exp\left({-2\gamma\, t}\right)$, where
$E_{+}(0)$ is also $E_1(0)/2$. Note that Eq.~\eqref{Q-}
is exact, as can be seen by replacing \eqref{Q+-} into \eqref{eqosc}.

The harmonic oscillators energies obtained from Eqs.~\eqref{lrt-q1}
and \eqref{lrt-q2} are compared in Fig.~\ref{fig4} with numerical
simulations of the Hamiltonian \eqref{eq1}. As seen from the f\/igure, the
qualitative agreement is excellent for both resonant (panel (a))\ and
non-resonant (panel (d))\ cases.

\begin{figure}[h!]
\begin{centering}
\includegraphics[scale=1]{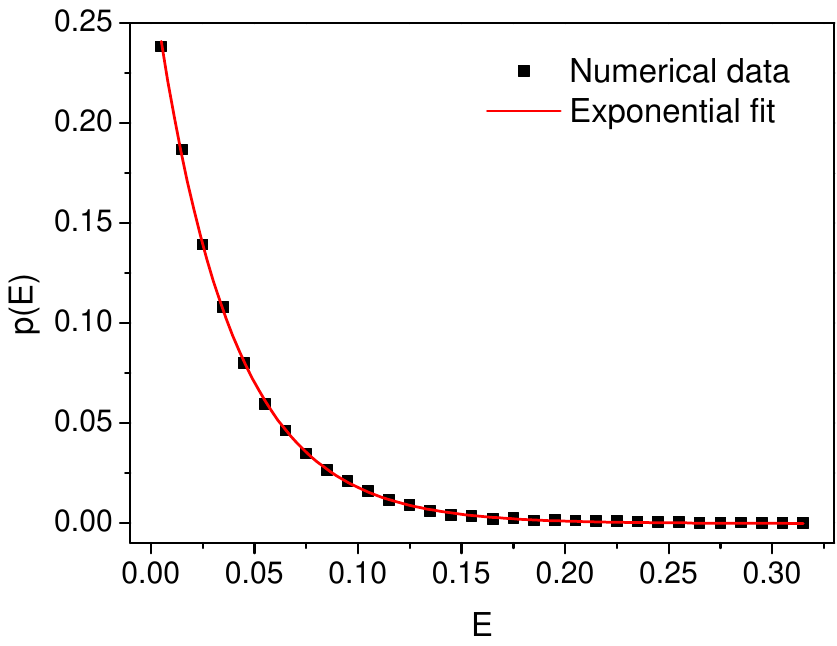}
\par
\end{centering}
\caption{(Color online)\ Distribution of energy for the dissipative
$Q_+$-mode for $N=100$. The f\/it is $p_B(E) = \exp{%
(-E/\Bar{E}_{+})} / \Bar{E}_{+}$ with $\Bar{E}_{+} = 0.037$.}
\label{fig6}
\end{figure}

In order to correctly apply the equipartition theorem in the resonant
case we have to consider only the energy in $Q_+$. Taking the expression
for the total initial energy, $E_T=E_{1}/2+NE_{QS}$, we f\/ind that the
expected value of the energy of the dissipative mode in equilibrium is
$\Bar{E}_{+} \approx 0.039$. This is conf\/irmed in Fig.~\ref{fig6},
which shows the energy distribution of $Q_+$. Therefore, the violation of
equipartition was only apparent, caused by the emergence of a conservative
mode that prevented the complete dissipation of the energy in the harmonic
oscillators.

\section{Conclusions}

We have studied the dynamical behavior of two harmonic oscillators
independently coupled to a chaotic environment with $2N$ degrees of
freedom. We focused our analysis on two points: the interactions
between the oscillators mediated by the chaotic environment and the
equilibrium properties of the system. We found that the oscillators can
exchange energy through the environment when in almost perfect
resonance. Deviations from this condition quickly changes this behavior
and makes the oscillators less sensitive to the presence of each other. When
in perfect resonance, the oscillators are able to keep part of their
initial energy in an apparent violation of the equipartion theorem.
This, however, turns out to be exactly the fraction of the energy
stored in the oscillators relative motion, which is not coupled to the
environment. This holds both for $N=1$ via ensemble averages or for
$N=100$ for a single realization of the dynamics and is also true
within the approximation of the linear response theory, which works
well for the resonant and non-resonant cases.

The equilibrium properties of the system, on the other hand, depend
critically on the number of degrees of freedom of the environment. In
order to quantify the equilibrium we considered two measures of
temperature: $\tau_E$, obtained from the equipartition theorem, and
$\tau_S$, obtained via the usual def\/inition of entropy. These
temperatures were also compared with the energy distribution of each
subsystem, whenever these converged to a Boltzmann-like exponential.
For $N=100$ the oscillator's energy distributions indeed always converged
to the exponential decay $\exp{(-E/\tau_B)}$ with $\tau_B \approx \tau_E$ for all
subsystems, except in the resonant cases. This characterizes the
thermal equilibrium and corroborates earlier results that a not too
small chaotic environment does play the role of a thermal bath
\cite{marchiori11}. This conclusion also holds in the resonant case if
the energy of the conservative mode is properly subtracted.

The case $N=1$, on the other hand, showed a very rich behavior.
Equilibration takes very long times for the present choice of
parameters and the energy distribution of the oscillators 
display, in some cases, exponential curves. Curiously, for the
particular values of $\omega_1$ and $\omega_2$ used, the exponent agreed
reasonably with energy equipartition, but not with $\tau_S$.
Moreover, the distribution of the momentum did often displayed the
expected Gaussian distribution, even if the corresponding energy
distribution was not exponential. These unexpected results show that the
$N=1$ environment does not simulate the action of a thermal bath even if
averaged over an ensemble of trajectories, and display a more complicated
type of approach to equilibrium that is worth a deeper investigation.

We have also examined the ergodicity of the system. Because $\lambda$ 
is small, it could be expected that the total system would not be ergodic. 
This, however, is not the case. For $\lambda=0$, the environment is 
ergodic for $N=1$ but it is not ergodic for $N=100$, since the quartic systems 
are totally independent from each other. As the coupling is turned on,
the total system becomes ergodic for $N=100$ (and is able to show the correct 
equipartition of energy), whereas, for $N = 1$, it does not. This was verif\/ied by 
numerical results not shown in which time and ensemble averages were 
compared for various conditions. As expected, we found that temperature
could be def\/ined whenever ergodicity was satisf\/ied, although ergodicity
itself depends non-trivially on the parameters of the system and on the number 
of degrees of freedom.

Finally, we note that the emergence of the conservative mode $Q_-$
depends on the symmetry of the coupling as given in
Eq.~\eqref{coupling} and on the resonance condition
$\omega_1=\omega_2$. The conservative mode is absent if
$\omega_1\neq\omega_2$ or if each oscillator is coupled to a
dif\/ferent mode of the quartic system, such as in $H_{I}=\sum_n(q_1
x_n + q_2 y_n)$. In LRT this leads to decoupled equations for $Q_+$ and
$Q_-$, each equation identical to Eq.~\eqref{new-oscillator}.

\begin{acknowledgments}
It is a pleasure to thank M.V.S.\ Bonan\c{c}a and T.F.\ Viscondi for helpful
suggestions. The authors acknowledge f\/inancial support from FAPESP
and CNPq, and computing facilities provided by CENAPAD/SP.
\end{acknowledgments}

\bibliographystyle{apsrev}

\end{document}